\documentclass[onecolumn]{IEEEtran}
\usepackage{rotating}
\usepackage{mathpazo}
\usepackage{times}
\usepackage{amsmath}
\usepackage{amsfonts}
\usepackage{latexsym}
\usepackage{amssymb}
\usepackage{cite}
\usepackage[ruled,vlined]{algorithm2e}
\usepackage{upref}
\usepackage{theorem}
\usepackage{graphicx}
\usepackage{multirow}
\usepackage{overpic}
\usepackage[usenames]{color}
 \usepackage {tikz}
\usetikzlibrary {positioning}
\usepackage{mathtools}
\usepackage{xparse}
\DeclarePairedDelimiterX{\set}[1]{\{}{\}}{\setargs{#1}}
\NewDocumentCommand{\setargs}{>{\SplitArgument{1}{;}}m}
{\setargsaux#1}
\NewDocumentCommand{\setargsaux}{mm}
{\IfNoValueTF{#2}{#1} {#1\,\delimsize|\,\mathopen{}#2}}%{#1\:;\:#2}

\DeclarePairedDelimiter\abs{\lvert}{\rvert}
\DeclarePairedDelimiter\ceil{\lceil}{\rceil}
\DeclarePairedDelimiter\floor{\lfloor}{\rfloor}
\DeclarePairedDelimiter\parenv{\lparen}{\rparen}

\DeclarePairedDelimiter\spn{\langle}{\rangle}

\theoremstyle{plain}
\newtheorem{theorem}{Theorem}
\newtheorem{lemma}[theorem]{Lemma}
\newtheorem{prop}[theorem]{Proposition}
\newtheorem{corollary}[theorem]{Corollary}
\newtheorem{definition}[theorem]{Definition}

\renewcommand{\leq}{\leqslant}

\renewcommand{\geq}{\geqslant}

\newcommand{\F}{\mathbb{F}}

\newcommand{\N}{\mathbb{N}}
\newcommand{\R}{\mathbb{R}}

\DeclareMathOperator{\wt}{wt}
\DeclareMathOperator{\supp}{supp}

\newcommand{\eqdef}{\triangleq}

\newcommand{\oc}{\overline{c}}

\newcommand{\bc}{\mathbf{c}}
\newcommand{\bs}{\mathbf{s}}

\newcommand{\ove}{\overline{e}}

\newcommand{\oh}{\overline{h}}
\newcommand{\bv}{\mathbf{v}}
\newcommand{\ov}{\overline{v}}
\newcommand{\bu}{\mathbf{u}}
\newcommand{\ou}{\overline{u}}

\newcommand{\os}{\overline{s}}
\newcommand{\ox}{\overline{x}}

\newcommand{\ozero}{\overline{0}}
\newcommand{\oone}{\overline{1}}

\newcommand{\T}{\intercal}

\DeclareMathOperator{\rmc}{RM}
\DeclareMathOperator{\rank}{rank}

\DeclareMathOperator*{\argmin}{argmin}

\begin{document}

%----------------- The Title Declarations ------------------------------

\title{On the Generalized Covering Radii \\ of Reed-Muller Codes}

\author{
  Dor Elimelech~\IEEEmembership{Student Member,~IEEE}, Hengjia Wei, and Moshe Schwartz~\IEEEmembership{Senior Member,~IEEE}%
  \thanks{Dor Elimelech is with the School
    of Electrical and Computer Engineering, Ben-Gurion University of the Negev,
    Beer Sheva 8410501, Israel
    (e-mail: doreli@post.bgu.ac.il).}%
  \thanks{Hengjia Wei is with the School
    of Electrical and Computer Engineering, Ben-Gurion University of the Negev,
    Beer Sheva 8410501, Israel
    (e-mail: hjwei05@gmail.com).}%
  \thanks{Moshe Schwartz is with the School
    of Electrical and Computer Engineering, Ben-Gurion University of the Negev,
    Beer Sheva 8410501, Israel
    (e-mail: schwartz@ee.bgu.ac.il).}%
  \thanks{The work of D. Elimelech was supported in part by an Israel Science Foundation (ISF) Grant under Grant 1052/18. The work of H. Wei and M. Schwartz was supported in part by a German Israeli Project Cooperation (DIP) Grant under Grant PE2398/1-1.}
} 

%% % make the title area
\maketitle

\begin{abstract}
We study generalized covering radii, a fundamental property of linear codes that characterizes the trade-off between storage, latency, and access in linear data-query protocols such as PIR. We prove lower and upper bounds on the generalized covering radii of Reed-Muller codes, as well as finding their exact value in certain extreme cases. With the application to linear data-query protocols in mind, we also construct a covering algorithm that gets as input a set of points in space, and find a corresponding set of codewords from the Reed-Muller code that are jointly not farther away from the input than the upper bound on the generalized covering radius of the code. We prove that the algorithm runs in time that is polynomial in the code parameters.
\end{abstract}


%%%%%%%%%%%%%%%%%%%%%%%%%%%%%%%%%%%%%%%%%%%%%%%%%%%%%%%%%%%%%%%
%%%%%%%%%%%%%%%%%%%%%%%%%%%%%%%%%%%%%%%%%%%%%%%%%%%%%%%%%%%%%%%
%%%%%%%%%%%%%%%%%%%%%%%%%%%%%%%%%%%%%%%%%%%%%%%%%%%%%%%%%%%%%%%

\section{Introduction}
\IEEEPARstart{T}{he} generalized covering radius has recently been proposed \cite{elimelech2020generalized} as a new fundamental property of linear codes, generalizing the classical notion  of a covering  radius. As a motivating application, these radii characterize a trade-off between storage, latency, and access complexities in linear data-query protocols, a prime example of which is the PIR (Private Information Retrieval) protocol. Several equivalent definitions of the generalized covering radii were given in \cite{elimelech2020generalized}, showing their combinatorial, geometric, and algebraic aspects. It has also been observed that there is an intriguing similarity between the generalized covering radii and the well known generalized Hamming weights of linear codes~\cite{1991-Wei}, hinting at a deeper theory and perhaps additional applications of these parameters that are yet to be revealed.

A crucial part in our understanding of any fundamental parameter of codes, is the values that it takes in specific examples and in parametric families of codes. In \cite{elimelech2020generalized}, the generalized covering radius hierarchy was found only for Hamming codes and shortened Hamming codes, whereas the remaining results did not pertain to specific code families. The Hamming code, in its extended version, is a specific case of the famous family of Reed-Muller codes, which is one of the most studied families of linear error-correcting codes. Reed-Muller codes have been extensively studied in the recent decades due to their practical applications and fascinating relations with various mathematical objects. Reed-Muller codes were recently proved to achieve asymptotically the capacity of erasure channels~\cite{kudekar2017reed}. They have long been conjectured to achieve Shannon’s capacity on symmetric channels, and a recent paper~\cite{abbe2020reed} took a step towards a proof of this conjecture, by showing a polarization property in Reed-Muller codes. Other applications of Reed-Muller codes include locally decodable code~\cite{yekhanin2012locally}, probabilistic proof systems~\cite{abbe2015reed}, sequence design for wireless communication~\cite{paterson2000generalized, davis1999peak, davis1997peak,schmidt2007complementary}, and Boolean functions~\cite{borissov2005covering, kurosawa2004new, meng2007analysis}. For a recent survey, the readers are referred to~\cite{abbe2021reed}.
 
While many aspects of Reed-Muller codes have been investigated, of particular interest to us is the (regular) covering radius. Its relation to the maximum nonlinearity of Boolean functions, motivated many of the papers on the subject. The covering radius of Reed-Muller codes has been studied in different settings~\cite{hou2006some, hou1993further, mcloughlin1979covering, cohen1992covering, carlet2006improving, helleseth1978covering, hou1997norm, schatz1981second, mykkeltveit1980covering, patterson1983covering}. However, despite decades of research on the subject, the exact covering radius of Reed-Muller codes is mostly unknown, except for a handful of specific cases, and many papers resorted to finding lower and upper bounds.

The goal of this paper is to explore the \emph{generalized covering radii} of Reed-Muller codes. Our main contributions are the following:
\begin{enumerate}
    \item 
    We prove lower and upper bounds on the generalized covering radii of Reed-Muller codes, $\rmc(r,m)$, in various asymptotic regimes of its parameters: constant $r$, constant $m-r$, and constant $r/m$. We also find the exact $t$-th generalized covering radius of $\rmc(r,m)$ in simple cases, $r\in \set*{0,m-2,m-1,m}$. These results are summarized in Table~\ref{tab:exact} and Table~\ref{tab:summary}.
    \item
    Motivated by the application for linear data-querying protocols, we construct a $t$-covering algorithm for Reed-Muller codes. Loosely speaking, given $t$ vectors in the space, the algorithm finds $t$ codewords that are jointly not farther away from the given points than the best upper bound that we have on the $t$-th generalized covering radius of the code. We analyze the run-time complexity of the algorithm and show it is polynomial in the code parameters.
\end{enumerate}

The paper is organized as follows: Preliminaries and notations are presented in Section~\ref{sec:prelim}. Section~\ref{sec:bounds} is devoted to the derivation of bounds on the generalized covering radii of Reed-Muller codes. The construction of our covering algorithm and its analysis are in Section~\ref{sec:alg}. We conclude with a discussion of the results and some open questions in Section~\ref{sec:conc}.

\section{Preliminaries}
\label{sec:prelim}

We use lower-case letters, $v$, to denote scalars, overlined lower-case letters, $\ov$, to denote vectors, and either bold lower-case letters, $\bv$, or upper-case letter, $V$, to denote matrices. Whether vectors are row vectors or column vectors is deduced from context.

Let $\F_q$ denote the finite field of size $q$. For $n\in\N$, we define $[n]\eqdef\set{1,\dots,n}$, and denote by $\binom{[n]}{t}$ the set of all subsets of $[n]$ of size $t$. For a vector $\ov=(v_1,\dots,v_n)\in \F_q^n$, the support of $\ov$ is defined as $\supp(\ov)\eqdef \set*{i\in[n] ; v_i\neq 0}$, and its Hamming weight is defined as $\wt(\ov)\eqdef \abs*{\supp(\ov)}$. The Hamming distance between $\ov,\ov'\in\F_q^n$ is then defined as $d(\ov,\ov')\eqdef\wt(\ov'-\ov)$.

We say $C$ is an $[n,k,d]_q$ linear code if $C\subseteq\F_q^n$ is a $k$-dimensional vector space, and the minimum Hamming distance between distinct codewords is $d$. The code $C$ may be specified using a $k\times n$ generator matrix $G\in\F_q^{k\times n}$, whose row space is $C$, or by an $(n-k)\times n$ parity-check matrix $H\in\F_q^{(n-k)\times n}$, whose null space is $C$. The dual code of $C$, denoted $C^\perp$, is the code whose generator matrix is $H$, and parity-check matrix is $G$, namely,
\[ C^\perp \eqdef \set*{ \ov\in\F_q^n ; \forall \oc\in C, \ov\cdot\oc=0}.\]
The dual code, $C^\perp$, is an $[n,n-k,d']_q$ code. We say $d'$ is the \emph{dual distance} of $C$.

For any vector $\ov\in\F_q^n$, the distance between $\ov$ and the code $C$ is defined as
\[ d(\ov,C)\eqdef \min_{\oc\in C}d(\oc,\ov).\]
The covering radius of $C$, denoted $R(C)$, is then defined as
\[ R(C) \eqdef \max_{\ov\in\F_q^n} d(\ov,C).\]
It is therefore the minimum radius at which balls centered at the codewords of $C$ cover the entire space $\F_q^n$. A generalization of this property will be presented shortly when we introduce the generalized covering radii of $C$. Later, we shall also make use of a connection between the covering radius of $C$, and the dual distance of $C$. To that end we recall the definition of Krawtchouk polynomials,
\[ K_k(x;n,q) \eqdef \sum_{j=0}^k (-1)^j \binom{x}{j}\binom{n-x}{k-j}(q-1)^{k-j},\]
where
\begin{equation}
    \label{eq:realbin}
\binom{x}{j}\eqdef \frac{x(x-1)\dots(x-j+1)}{j!}.
\end{equation}
We further denote the minimal root of $K_k(x;n,q)$ by
\[ x(k,n;q) \eqdef \min\set*{x\in\R ; K_k(x{;}\, n,q)=0}.\]

\begin{lemma}
\label{lem:dualdist}
\cite[Theorem~3.3]{Tietavainen91}
Let $C$ be an $[n,k]_q$ code with dual distance $d'$. Then
\[ R(C)\leq \begin{cases}
x(u,n-1;q) & d'=2u-1,\\
x(u,n;q) & d'=2u.
\end{cases}
\]
\end{lemma}
    
\subsection{The generalized covering radii}

The generalized covering radii of a linear code were introduced in \cite{elimelech2020generalized}. They have several equivalent definitions, which we bring here and use interchangeably. We begin with a geometric definition. Consider the set of matrices $\F_q^{t \times n}$, in which we have a generalized notion for the Hamming weights. For a matrix $\bv\in \F_q^{t\times n}$, with row vectors denoted by $\ov_i$, the $t$-weight is defined to be
\[  \wt^{(t)}(\bv)\eqdef \abs*{\bigcup_{i\in [t]} \supp(\ov_i)} .\] 
The $t$-weight naturally induces a metric on  $ \F_q^{t\times n}$ by 
 \[ d^{(t)}(\bv,\bu)\eqdef \wt^{(t)}(\bv-\bu), \]
for all $\bv,\bu\in \F_q^{t\times n}$. Let $B_r^{(t)}(\bv)$ denote the ball of radius $r$ centered in $\bv\in \F_q^{t\times n}$, with respect to the metric $d^{(t)}$, namely
\[ B_r^{(t)}(\bv) \eqdef\set*{ \bv'\in\F_q^{t\times n} ; d^{(t)}(\bv,\bv')\leq r}.\]
Since this metric is translation invariant, the volume of the ball does not depend on the choice of its center. We denote this volume by
\begin{equation}
\label{eq:ballvol}
V_{q^t,n,r}\eqdef \abs*{B_r^{(t)}(\bv)}=\sum_{i=0}^r \binom{n}{i}(q^t-1)^i,
\end{equation}
which is exactly the size of a ball of radius $r$ in $\F_{q^t}^n$ using the Hamming metric. We now have the following definition for the $t$-th generalized radius:
\begin{definition}
\label{def:rt2}
Let $C$ be an $[n,k]_q$ linear code. Then for every $t\in\N$, we define the $t$-th generalized covering radius, denoted by $R_t(C)$, to be the minimal integer $r$ such that the balls of radius $r$ centered at
\begin{equation}
    \label{eq:tpow}
C^t \eqdef \set*{ \begin{bmatrix} \oc_1 \\ \vdots \\ \oc_t\end{bmatrix} \in 
\F_q^{t\times n}; \forall i\in[t], \oc_i\in C},
\end{equation}
cover $\F_q^{t\times n}$, i.e.,
\[\bigcup_{\bc\in C^t}B^{(t)}_r(\bc) = \F_q^{t\times n}.\]
\end{definition}

One can easily see that $R_1(C)=R(C)$ is indeed the regular covering radius of the code $C$. Let us now turn to an equivalent definition via the parity-check matrix of a code. Assume $C$ is a linear $[n,k]_q$ code with a (full-rank) parity-check matrix $H\in\F_q^{(n-k)\times n}$. Let the columns of $H$ be denoted by $\oh_1,\dots,\oh_n$. Then for $I\in \binom{[n]}{t}$, $1\leq t\leq n$, we denote the linear span of $\set{\oh_i}_{i\in I}$ by $\spn{H_I}$. We have the following equivalent definition for the $t$-th generalized covering radius of $C$:

\begin{definition}
\label{def:rt1}
The $t$-th covering radius of $C$, denoted by $R_t(C)$, is the smallest integer $r$ such that for any $t$ vectors $\ov_1,\dots, \ov_t\in \F_q^{n-k}$, there exists $I\in\binom{[n]}{r}$ such that $\set{v_1,\dots,v_t}\subseteq \spn{H_I}$.
\end{definition}

The final equivalent definition that we recall for the generalized covering
radius is algebraic in nature:

\begin{definition}
\label{def:rt3}
Let $C\subseteq \F_q^n$ be a linear code with a generator matrix $G\in\F_q^{k\times n}$ and a parity-check matrix $H\in\F_q^{(n-k)\times n}$. Let $C_t$ be the code over $\F_{q^t}$, with generator matrix $G$ and parity-check matrix $H$, namely,
\begin{equation}
    \label{eq:ct}
    C_t\eqdef \set*{ \ou G ; \ou\in\F_{q^t}^k} = \set*{ \ov\in\F_{q^t}^n ; H\oc^\T = \ozero^\T}.
\end{equation}
The $t$-th covering radius is defined to be \[ R_t(C)\eqdef R_1(C_t),\]
where $R_1(C_t)$ is the (regular, first) covering radius of $C_t$.
\end{definition}

According to Definition~\ref{def:rt3}, the problem of finding the $t$-th covering radius of a code $C\subseteq \F_q^n$, is equivalent to finding the regular covering radius of $C_t$ defined over $\F_{q^t}$. Since the code $C_t$ will be used many times, we briefly show that, unlike the covering radius, its minimum distance does not change.

\begin{lemma}
\label{lem:ctdist}
Let $C$ be an $[n,k,d]_q$ code. Then for any $t\in\N$, the code $C_t$ of~\eqref{eq:ct} is an $[n,k,d]_{q^t}$ code.
\end{lemma}
\begin{IEEEproof}
The fact that $C_t$ has length $n$ is trivial. Let $H\in\F_q^{(n-k)\times n}$ be parity-check matrix for $C$. Since a set of vectors from $\F_q^n$ is linearly independent over $\F_q$ if and only if it is linearly independent over $\F_{q^t}$, the matrix $H$ has the same rank over $\F_{q^t}$, and its null-space, $C_t$, has dimension $k$. Finally, it is well known that the minimum distance $d$ of $C$ is the minimal number of columns of $H$ that are linearly dependent. By the same argument as before, this number does not change when considering columns of $H$ and linear dependence over $\F_{q^t}$. Hence, the minimum distance of $C_t$ is also $d$.
\end{IEEEproof}

The generalized covering radius has a subadditivity property that proves to be useful for establishing many of the results in this work:

\begin{lemma}
\label{lem:subadd}
\cite[Proposition 15]{elimelech2020generalized}
Let $C$ be an $[n,k]_q$ code. Then for all $t_1,t_2\in\N$,
\[ R_{t_1+t_2}(C) \leq R_{t_1}+R_{t_2}(C).\]
In particular, $R_{t}(C) \leq t R_{1}(C)$ for all $t\in \N$.
\end{lemma}

A simple ball-covering argument is used in the following lemma.

\begin{lemma}
\label{lem:SphereBoundGeneral}
For an $[n,k]_q$ code $C$ and $t\in \N$, 
\[ \log_{q^t} \parenv*{V_{q^t,n,R_t(C)}} \geq n-k. \] 
\end{lemma}

\begin{IEEEproof}
Recalling~\eqref{eq:ct}, consider the code $C_t$ over $\F_{q^t}$,  generated by the same generator matrix as $C$. Clearly, $C_t$ has the same dimension and length as $C$. By the standard ball-covering argument (see \cite[Theorem 6.2.1]{cohen1997covering}), 
\[ \log_{q^t} \parenv*{V_{q^t,n,R_1(C_t)}}\geq n-k. \]
By Definition~\ref{def:rt3}, $R_1(C_t)=R_t(C)$, and we conclude. 
\end{IEEEproof}

Since we shall be interested in asymptotic results, we recall facts about the asymptotics of binomial coefficients as well as the volume of balls in the Hamming metric. Let $H_q(x)$ denote the $q$-ary entropy function,
\[ H_q(x)\eqdef x\log_q(q-1) - x\log_q(x)-(1-x)\log_q (1-x).\]
A useful Taylor expansion near the entropy function's maximum was presented in~\cite[Proposition 3.3.5]{Guruswamietal2019}, showing that, as $\epsilon\to 0$,
\begin{equation}
    \label{eq:enttaylor}
H_q\parenv*{1-\frac{1}{q}-\epsilon}=1-\frac{\epsilon^2 q^2}{2(q-1)\ln q}(1+o(1)).
\end{equation}
For any real $0<\alpha<1$, such that $\alpha n\in\N$, it is known (e.g., see~\cite[Ch.~10, Lemma 7]{macwilliams1977theory}) that
\begin{equation}
    \label{eq:asymbinom}
    \frac{1}{\sqrt{8n\alpha(1-\alpha)}}2^{nH_2(\alpha)}\leq \binom{n}{\alpha n} \leq \frac{1}{\sqrt{2\pi n\alpha(1-\alpha)}}2^{nH_2(\alpha)},
\end{equation}
and this holds for $n\in\R$, $n>1$ (recall the definition of the binomial in~\eqref{eq:realbin}, and see~\cite[p.~482]{GraKnuPat94}). As for the Hamming ball, it is well known (see~\cite[Ch.~10, Corollary 9]{macwilliams1977theory} and~\cite[Proposition~3.3.1]{Guruswamietal2019}) that for $q\geq 2$, and $\alpha\leq 1-\frac{1}{q}$,
\begin{equation}
    \label{eq:asymballvol}
    \frac{1}{\sqrt{8n\alpha(1-\alpha)}}q^{nH_q(\alpha)} \leq V_{q,n,\alpha n} \leq q^{nH_q(\alpha)}.
\end{equation}

\subsection{Reed-Muller codes}

Reed-Muller codes have been extensively studied (e.g., see~\cite{macwilliams1977theory}, and the many references therein). We recall the relevant definitions and properties needed for this paper. For $m\in \N$ and $0\leq r \leq m$, the $r$-th order Reed-Muller code, denoted by $\rmc(r,m)$, is a binary linear $[n,k]$ code with parameters
\begin{align}
\label{eq:rmparam}
    n&=2^m, & k&=\sum_{i=0}^r \binom{m}{i}.
\end{align}
Reed-Muller codes have multiple equivalent definitions, and one that will be useful for our needs is a recursive definition, given by the $(u,u+v)$ construction. 
Assume $C_1$ and $C_2$ are $[n,k_1]_q$ and $[n,k_2]_q$ codes, respectively.
The $(u,u+v)$ construction uses $C_1$ and $C_2$ to produce a code
\[ C=\set*{(\ou,\ou+\ov) ; \ou\in C_1,\ov\in C_2}.\]
As a base for the recursion, we define 
\[\rmc(0,m)\eqdef\set*{\ozero,\oone},\]
i.e., the repetition code. Additionally, we define
\[\rmc(m,m)\eqdef\F_2^{2^m},\]
i.e., the entire set of binary vectors of length $2^m$. Finally, for $1\leq r \leq m-1$, we define $\rmc(r,m)$ to be the code produced by the $(u,u+v)$ construction using $\rmc(r,m-1)$ and $\rmc(r-1,m-1)$.

Reed-Muller codes are nested, namely, for all $1\leq r\leq m$,
\begin{equation}
    \label{eq:nested}
\rmc(r-1,m)\subseteq \rmc(r,m).
\end{equation}
Additionally, the family of Reed-Muller code is closed under code duality, and in particular
\[ \rmc(r,m)^\perp = \rmc(m-r-1,m).\]
This implies that
\begin{equation}
    \label{eq:dimdual}
\dim\parenv*{\rmc(r,m)}=2^m-\dim\parenv*{\rmc(m-r-1,m)}.
\end{equation}

To avoid cumbersome notation, we denote the $t$-th generalized covering radius of the $r$-th order Reed-Muller code by
\[R_t(r,m)\eqdef R_t(\rmc(r,m)).\]
The following fundamental property of $R_t(r,m)$ will be used frequently in this work:

\begin{prop}
\label{prop:recursive}
For all $m,t\in \N$, and $1\leq r\leq m-1$,
\[R_t(r,m)\leq R_t(r-1,m-1)+R_t(r,m-1).\]
\end{prop}
\begin{IEEEproof}
The claim follows from the $(u,u+v)$ construction of Reed-Muller codes. In \cite[Proposition 24]{elimelech2020generalized} it is proved that if a code $C$ is produced using the $(u,u+v)$ construction with $C_1$ and $C_2$, then $R_t(C)\leq R_t(C_1)+R_t(C_2)$.
\end{IEEEproof}

\section{Bounds}
\label{sec:bounds}
Our main results are presented in this section. We prove bounds on the generalized covering radii of Reed-Muller codes, $\rmc(r,m)$, in three different asymptotic regimes, as $m\to\infty$:
\begin{itemize}
    \item  $r$ is constant.
    \item  $m-r$ is constant.
    \item  $r/m$ is constant.
\end{itemize}
Upper bounds will be derived using two main strategies: The first is by considering the upper bounds from \cite{cohen1992covering} and using the subadditivity formula from Lemma~\ref{lem:subadd}. The second strategy involves the use of the recursive formula from Proposition~\ref{prop:recursive} and analysis of the base cases. Our lower bounds will essentially be the well known ball-covering lower bound (over the field $\F_{q^t}$), analyzed separately for each of the different cases.

\subsection{The case where $r$ is constant}

In this parameter regime, the Reed-Muller codes have vanishing asymptotic rate, and high covering radius. We first consider the extreme case of $\rmc(0,m)$, which is none other than the repetition code. In this simple case we can determine the generalized covering radii exactly.

\begin{prop}
\label{prop:rtzerom}
For all $m,t\in\N$,
\[ R_t(0,m) = 2^m-\ceil*{2^{m-t}}.\]
\end{prop}

\begin{IEEEproof}
The Reed-Muller code $C=\rmc(0,m)$ is the binary repetition code of length $2^m$, namely, its generator matrix is $G=(1,1,\dots,1)$. Thus, $C_t$ of~\eqref{eq:ct} is just the $2^t$-ary repetition code of the same length. Given a vector $\ov\in \F_{2^t}^{2^m}$, the closest codeword of $C_t$ to $\ov$ is $\oc=(c,c,\dots,c)\in C_t$ where $c\in\F_{2^t}$ is the symbol appearing the most times in $\ov$. By simple averaging, there exists a symbol appearing at least $\ceil{2^{m-t}}$ times in $\ov$, giving us $R_t(0,m)\leq 2^m-\ceil{2^{m-t}}$. For the lower bound, define $\ell\eqdef\min\set{t,m}$, and let $\ov\in \F_{2^t}^{2^m}$ be a vector with $2^\ell$ different symbols, such that each symbol appears exactly $2^{m-\ell}$ times. Clearly, we have 
\[d(\ov,\rmc(0,m))=2^m - 2^{m-\ell} \geq 2^m -\ceil{2^{m-t}}. \] 
This proves  the lower bound.
\end{IEEEproof}

For the more general cases of $\rmc(r,m)$ with $r\geq 1$, we provide separate upper and lower bound on the generalized covering radii. The upper bounds are proved by induction on $r$. The base case of $\rmc(1,m)$ is proved first.

\begin{lemma}\label{lm:upperbound-s=1}
For all $m,t\in\N$,
\[R_t(1,m) \leq \parenv*{1-\frac{1}{2^t}}2^m-\frac{\sqrt{2^t-1}}{2^t}2^{m/2}.\]
\end{lemma}

\begin{IEEEproof}
Denote $C=\rmc(1,m)$. It is well known that $C^\perp=\rmc(m-2,m)$ is the extended binary Hamming code (see~\cite[Ch.~13]{macwilliams1977theory}), and hence the dual distance of $C$ is $d'=4$. By Lemma~\ref{lem:ctdist}, $d'=4$ is the dual distance of $C_t$ of~\eqref{eq:ct} as well. By Lemma~\ref{lem:dualdist}, the covering radius of $C_t$ is upper bounded by
\[R_t(C)=R_1(C_t) 
\leq x(2,2^m;2^t),\]
i.e., the smallest root of the Krawtchouk polynomial $K_2(x;2^m,2^t)$. Since
\[K_2(x;q,n)=\frac{1}{2}\parenv*{ q^2x^2-q(2qn-q-2n+2)x+(q-1)^2 n(n-1) },\]
it follows that 
\begin{align*} 
x(2,n;q)=\parenv*{1-\frac{1}{q}}n-\frac{1}{2}+\frac{1}{q}-\frac{\sqrt{(4q-4)n+(q-2)^2}}{2q} \leq \parenv*{1-\frac{1}{q}}n-\frac{\sqrt{(q-1)n}}{q}.
\end{align*}
Plugging in $n=2^m$ and $q=2^t$, we obtain the desired result.
\end{IEEEproof}

 We can now prove the general upper bound on $R_t(r,m)$ for $r\geq 1$.

\begin{theorem}\label{th:upperbound-fixs}
For all $m,t\in\N$, $1\leq r\leq m$,
\[
    R_t(r,m) \leq \parenv*{1-\frac{1}{2^t}}2^m-\frac{\sqrt{2^t-1}}{2^t} (1+\sqrt{2})^{r-1}2^{m/2}+O(m^{r-2}).
\]
where we consider $r$ and $t$ to be constants.
\end{theorem}

\begin{IEEEproof}
We prove the claim by induction on $r$. Lemma~\ref{lm:upperbound-s=1} shows the claim holds for $r=1$, and for all $m\in\N$. Assume that the claim holds for all $\ell\leq r-1$, and all $m\in\N$. We now show that it holds for $r$ as well. By repeatedly using Proposition~\ref{prop:recursive} and the induction hypothesis, we have,
\begin{align*}
    R_t(r,m) & \leq R_t(r,m-1) + R_t(r-1,m-1) \\
     & \leq R_t(r,m-1) + \parenv*{1-\frac{1}{2^t}}2^{m-1}-\frac{\sqrt{2^t-1}}{2^t} (1+\sqrt{2})^{r-2}2^{(m-1)/2}+O(m^{r-3})\\
     & \vdots \\
     & \leq R_t(r,r)+ \sum_{i=r}^{m-1}\parenv*{\parenv*{1-\frac{1}{2^t}} 2^i -\frac{\sqrt{2^t-1}}{2^t} (1+\sqrt{2})^{r-2} 2^{i/2}+O(m^{r-3})} \\
     & \leq R_t(r,r)+ \parenv*{1-\frac{1}{2^t}} \sum_{i=0}^{m-1}2^i -\frac{\sqrt{2^t-1}}{2^t} (1+\sqrt{2})^{r-2} \sum_{i=r}^{m-1}2^{i/2}+O(m^{r-2}) \\
     & = \parenv*{1-\frac{1}{2^t}}2^m-\frac{\sqrt{2^t-1}}{2^t} (1+\sqrt{2})^{r-1}\parenv*{2^{m/2}-2^{r/2}}+O(m^{r-2})
     \\
     & = \parenv*{1-\frac{1}{2^t}}2^m-\frac{\sqrt{2^t-1}}{2^t} (1+\sqrt{2})^{r-1}2^{m/2}+O(m^{r-2}).
\end{align*}
Here we also use the fact $R_t(r,r)=0$, since $\rmc(r,r)=\F_2^{2^r}$, and so $\rmc(r,r)_t=\F_{2^t}^{2^r}$, whose covering radius is $0$.
\end{IEEEproof}

The corresponding lower bound on $R_t(r,m)$ is proved next. It is obtained by carefully considering both a ball-covering argument, and the upper bound we just proved.

\begin{theorem}\label{th:lowerbound-fixs}
For all $m,t\in\N$, $1\leq r\leq m$,
\begin{equation}\label{eq:lowerbound-fixs}
    R_t(r,m) \geq \parenv*{1-\frac{1}{2^t}}2^m-\frac{\sqrt{2t(2^t-1) \ln 2}}{2^t\sqrt{r!}} m^{r/2}2^{m/2}(1+o(1)),
\end{equation}
where we consider $r$ and $t$ to be constants.
\end{theorem}

\begin{IEEEproof}
By Lemma~\ref{lem:SphereBoundGeneral}, we have that 
\[ \log_{2^t} \parenv*{V_{2^t,2^m,R_t(r,m)}} \geq 2^m -\sum_{i=0}^r \binom{m}{i}=2^m-\frac{m^r}{r!}(1+o(1)). \]
According to Theorem~\ref{th:upperbound-fixs},
\begin{equation}
    \label{eq:defy}
\frac{R_t(r,m)}{2^m}=1-\frac{1}{2^t}-o(1),
\end{equation}
and in particular, for all large enough $m$, 
\[\frac{R_t(r,m)}{2^m}<1-\frac{1}{2^t}.\]
Using~\eqref{eq:asymballvol},
\[ \log_{2^t} \parenv*{V_{2^t,2^m,R_t(r,m)}} \leq 2^m H_{2^t} \parenv*{\frac{R_t(r,m)}{2^m}}.\]
Combining the two inequalities above, we have 
\begin{equation}\label{eq:coveringbound}
    2^m H_{2^t} \parenv*{\frac{R_t(r,m)}{2^m}} \geq 2^m-\frac{m^r}{r!}(1+o(1)).
\end{equation}

Denote $y\eqdef 1-1/2^t-R_t(r,m)/2^m$. Then $y=o(1)$ by~\eqref{eq:defy}, and $y>0$ for all large enough $m$. Thus, by~\eqref{eq:enttaylor} we have
\[ H_{2^t} \parenv*{\frac{R_t(r,m)}{2^m}} =   H_{2^t} \parenv*{1-\frac{1}{2^t}-y} = 1-cy^2(1+o(1)),\]
where $c=\frac{2^{2t}}{2t(2^t-1)\ln 2}$. Hence,
\[1-cy^2 (1+o(1))\geq 1 - \frac{m^r}{2^m r!} (1+o(1)),\]
and so,
\[y \leq \frac{m^{r/2}}{2^{m/2} \sqrt{r!\cdot c}} (1+o(1)). \]
The conclusion follows since $R_t(r,m)=(1-1/2^t-y)2^m$.
\end{IEEEproof}

\subsection{The case where $m-r$ is constant}

The opposite case to the one studied in the previous section, is that of Reed-Muller codes $\rmc(r,m)$ with $m-r$ being constant. These codes have a high rate and a vanishing normalized covering radius. As we show shortly, in this asymptotic regime, the $t$-th generalized covering radius is approximately linear in $t$. We begin, however, with the two extreme cases of $\rmc(m-1,m)$ and $\rmc(m-2,m)$.

\begin{prop}
\label{prop:rtalmostm}
For all $m,t\in\N$,
\begin{align*}
    R_t(m,m) & = 0, \\
    R_t(m-1,m) & = 1, \\
    R_t(m-2,m) & = \min\set*{t,m}+1.
\end{align*}
\end{prop}

\begin{IEEEproof}
The case of $R_t(m,m)$ is trivial since $\rmc(m,m)=\F_2^{2^m}$. For the next case, $\rmc(m-1,m)$ is the binary $[2^m,2^m-1,2]$ parity code. Its parity-check matrix is $H_1=(1,1,\dots,1)$. Then, by directly using Definition~\ref{def:rt1}, we get that for all $t\in\N$, $R_t(m-1,m)=1$.

Finally, $\rmc(m-2,m)$ is the binary $[2^m,2^m-m-1,4]$ extended Hamming code. A parity-check matrix for this code is the $(m+1)\times 2^m$ matrix $H_2$ containing all the binary column vectors that start with a $1$. Let $\ove_i$ denote the $i$-th standard unit column vector. We again use Definition~\ref{def:rt1} directly: for any $1\leq t\leq m$, we contend that the set $\set{\ove_2,\ove_3,\dots,\ove_{t+1}}$ cannot be spanned by $t$ columns of $H_2$. That is because $\spn{\ove_2,\dots,\ove_{t+1}}$ is a $t$-dimensional vector space, all of whose vectors contain a $0$ in the first coordinate. However, the span of any $t$ columns from $H_2$ is, at best, a $t$-dimensional vector space, but whose vectors' first coordinate is not always $0$. Thus, $R_t(m-2,m)\geq t+1$. However, given any set of $t$ column vectors of length $m+1$, $\set*{\ov_1,\dots,\ov_t}$, the set is spanned by the  $t+1$ vectors $\set*{\ov_1',\ov_2',\dots,\ov_t',\ove_1}$ where $\ov_i'=\ov_i$ if the first coordinate of $\ov_i$ is $1$ and $\ov_i'=\ov_i+\ove_1$ otherwise.  Clearly,  $\set*{\ov_1',\ov_2',\dots,\ov_t',\ove_1}$ are all columns of $H$, and therefore, $R_t(m-2,m)\leq t+1$. Combining the two bounds we get that $R_t(m-2,m)=t+1$, for all $t\leq m$. Finally, for $t>m$ the claim is trivial since $\rank(H_2)=m+1$, and any set of column vectors of length $m+1$ can be spanned by $m+1$ linearly independent columns of $H_2$.
\end{IEEEproof}

Turning to the more general case of $\rmc(m-s,m)$, we first prove a technical lemma. The proof of this lemma is primarily based on the estimation of binomial coefficients by Stirling's approximation.

\begin{lemma}
\label{lem:BallEstimate}
Let  $t\in \N$ be a constant, and $r=o\parenv*{2^m}$. Then
\[ \log_{2^t}\parenv*{V_{2^t,2^m,r}}=\frac{mr}{t} - O(r\log(r)).\] 
\end{lemma}

\begin{IEEEproof}
Since $r=o\parenv*{2^m}$, for sufficiently large $m$ we have that $r< 2^{m-1}$, and  therefore 
\[\binom{2^m}{i}(2^t-1)^i\leq \binom{2^m}{i+1}(2^t-1)^{i+1},\]
for all $0\leq i \leq r$. It follows that
\begin{align}
\label{eq:binbound}
    \binom{2^m}{r} 2^{r(t-1)} \leq  V_{2^t,2^m,r}=\sum_{i=0}^r  \binom{2^m}{i}(2^t-1)^i \leq (r+1) \binom{2^m}{r} 2^{rt}.
\end{align}
By Stirling's approximation (e.g., see~\cite[p.~251]{GraKnuPat94}),
\[ \parenv*{\frac{2^m}{r}}^r\leq \binom{2^m}{r}\leq \parenv*{e\frac{2^m}{r}}^r.\] Applying $\log_{2^t}$ and simplifying we obtain,
\begin{align}
    \label{eq:stirling}
    \frac {mr}{t}-r\log_{2^t}\parenv*{r} \leq \log_{2^t}\binom{2^m}{r}\leq \frac{mr}{t}-r\log_{2^t}\parenv*{\frac{r}{e}}.
\end{align}
Combining~\eqref{eq:binbound} and~\eqref{eq:stirling} we have 
\begin{align*}
     \log_{2^t} \parenv*{V_{2^t,2^m,r}} &\leq \log_{2^t} \parenv*{ (r+1) \binom{2^m}{r} 2^{rt}}\leq \log_{2^t}(r+1) + \frac {mr}{t}-r\log_{2^t}\parenv*{\frac{r}{e}}+r\\
     &=  \frac {mr}{t}-O(r\log(r)).
\end{align*}
Similarly, 
\begin{align*}
     \log_{2^t} \parenv*{V_{2^t,2^m,r}} &\geq \log_{2^t} \parenv*{  \binom{2^m}{r} 2^{r(t-1)}}\geq \frac{mr}{t}-r\log_{2^t}\parenv*{r}+\frac{r(t-1)}{t}\\
     &=  \frac {mr}{t}-O(r\log(r)).
\end{align*}
\end{IEEEproof}

We can now state the main bounds for this asymptotic regime.

\begin{theorem}
\label{th:LB2}
For all $m,t\in\N$, $3\leq s\leq m$,
\[ \frac{t}{(s-1)!} m^{s-2}+O(m^{s-3}\log(m)) \leq R_t(m-s,m)\leq \frac{t}{(s-2)!}m^{s-2}+O(m^{s-3}),\] 
where we consider $s$ and $t$ to be constants.
\end{theorem}

\begin{IEEEproof}
In \cite[Section 3]{cohen1992covering} it is proved for the (first) covering radius that 
\[R_1(m-s,m)\leq \frac{m^{s-2}}{(s-2)!}+O(m^{s-3}).\]
Combining this with Lemma~\ref{lem:subadd}, the upper bound follows immediately.

Having proven the upper bound, we see that $R_t(m-s,m)=o(2^m)$. Thus, by Lemma~\ref{lem:BallEstimate},
\[ \log_{2_t} \parenv*{V_{2^t,2^m,R_t(m-s,m)}}=\frac{mR_t(m-s,m)}{t}-O(m^{s-2}\log(m)). \] 
Combining this with the ball-covering argument from Lemma~\ref{lem:SphereBoundGeneral} and~\eqref{eq:dimdual}, it follows that
\begin{align*}
\frac{mR_t(m-s,m)}{t}-O(m^{s-2}\log(m)) &=\log_{2_t} \parenv*{V_{2^t,2^m,R_t(m-s,m)}}\geq 2^m-\dim\parenv*{\rmc(m-s,m)}\\
&=\dim\parenv*{\rmc(s-1,m)}
= \sum_{i=0}^{s-1}\binom{m}{i} \geq \binom{m}{s-1} \geq \frac{m^{s-1}}{(s-1)!}-O(m^{s-2}).
\end{align*}
After rearranging we get the claim.
\end{IEEEproof}

\subsection{The case where $r/m$ is constant }

The final asymptotic regime we study is when $r/m=\alpha$ is constant. For technical reasons, we divide the discussion into two different cases: $\frac{1}{2}<\alpha<1$, and $0<\alpha\leq\frac{1}{2}$. We begin with the range $\frac{1}{2}<\alpha<1$.

\begin{theorem}
\label{th:highalpha}
For all $m,t\in\N$ and $\frac{1}{2}<\alpha<1$,
\[
t\cdot \sqrt{\frac{1-\alpha}{8(\alpha m)^3}}\cdot  2^{mH_2(\alpha)}\cdot(1+o(1)) \leq R_t(\alpha m,m)\leq t\cdot 4^{H_2(\alpha)}\cdot 2^{mH_2(\alpha)}\cdot (1+o(1)),
\]
where we consider $t$ and $\alpha$ to be constants.
\end{theorem}
\begin{IEEEproof}
In \cite[Theorem 9.4.25]{cohen1997covering} it is proved that for $\frac{1}{2}< \alpha<1$, the (first) covering radius satisfies
\[ R_1(\alpha m,m)\leq 4^{H_2(\alpha)}\cdot 2^{mH_2(\alpha)}\cdot (1+o(1)). \]
By applying the subadditivity property from Lemma~\ref{lem:subadd} we immediately obtain the claimed upper bound.

For the lower bound, as in the proof of Theorem~\ref{th:LB2},
\begin{align*}
\log_{2^t}\parenv*{V_{2^t,2^m,R_t(\alpha m,m)}} & \geq \dim\parenv*{\rmc((1-\alpha)m-1,m)} = \sum_{i=0}^{(1-\alpha)m-1 }\binom{m}{i} \\
& \geq \binom{m}{(1-\alpha)m-1} = \frac{(1-\alpha)m}{\alpha m+1}\binom{m}{(1-\alpha)m}=\frac{1-\alpha}{\alpha}\binom{m}{(1-\alpha)m}(1+o(1))\\
& \geq \sqrt{\frac{1-\alpha}{8m\alpha^3}}\cdot  2^{mH_2(\alpha)}\cdot(1+o(1)),
\end{align*}
where the last inequality follows from~\eqref{eq:asymbinom}. By the Upper bound presented above, $R_t(\alpha m,m)=o(2^m)$, and Lemma~\ref{lem:BallEstimate} may be applied to obtain
\[ \frac{m R_t(\alpha m,m)}{t}(1+o(1))=\log_{2^t}(V_{2^t,2^m,R_t(\alpha m,m)})\geq \sqrt{\frac{1-\alpha}{8m\alpha^3}}\cdot  2^{mH_2(\alpha)}\cdot(1+o(1)) .\]
By rearranging we obtain the desired lower bound.
\end{IEEEproof}

We now move on to the range $0<\alpha\leq\frac{1}{2}$. We begin with two lemmas, laying the groundwork for the bounds. The first lemma is a weaker, more general version of an upper bound on $R_t(r,m)$.

\begin{lemma}
\label{lem:BinomBound}
For all $m,t\in\N$, $1\leq r\leq m$,
\[ R_t(r,m)\leq \parenv*{1-\frac{1}{2^t}}2^m -\frac{\sqrt{2^t-1}}{2^t}\binom{m}{r}.\] 
\end{lemma}
\begin{IEEEproof}
We prove the claim by induction on $m$. We first observe that the the claim holds in the extreme cases where $r=1$ and $r=m$. Since $2^{m/2}\geq m=\binom{m}{1}$ for any $m\in\N$, by Lemma~\ref{lm:upperbound-s=1} we have
\[ R_t(1,m)\leq \parenv*{1-\frac{1}{2^t}}2^m -\frac{\sqrt{2^t-1}}{2^t} 2^{\frac{m}{2}}\leq \parenv*{1-\frac{1}{2^t}}2^m -\frac{\sqrt{2^t-1}}{2^t} \binom{m}{1}.  \] 
In the case where $r=m$, $\rmc(m,m)=\F_2^{2^m}$, and thus $R_t(m,m)=0$ and the claim holds. In particular, this proves the claim for $m=1,2$, serving as the induction base.

Assume the claim holds for $m-1$, and we now prove it holds for $m$. We already know the claim holds for $R_t(1,m)$ and $R_t(m,m)$. Thus, we only need to show it holds for $2\leq r\leq m-1$. By Proposition~\ref{prop:recursive} and the induction hypothesis,
\begin{align*}
     R_t(r,m)& \leq  R_t(r-1,m-1)+R_t(r,m-1) \\
     &\leq \parenv*{1-\frac{1}{2^t}}2^{m-1} -\frac{\sqrt{2^t-1}}{2^t}\binom{m-1}{r-1}+ \parenv*{1-\frac{1}{2^t}}2^{m-1} -\frac{\sqrt{2^t-1}}{2^t}\binom{m-1}{r}\\
     &=\parenv*{1-\frac{1}{2^t}}2^{m} -\frac{\sqrt{2^t-1}}{2^t}\binom{m}{r},
\end{align*}
thus completing the induction step.
\end{IEEEproof}

The next technical lemma proves the limit of $R_t(\alpha m,m)/2^m$.

\begin{lemma}
\label{lem:limitrtlowalpha}
Let $0<\alpha\leq\frac{1}{2}$ be a constant. Then
\[ \lim_{m\to\infty} \frac{R_t(\alpha m,m)}{2^m}= 1-\frac{1}{2^t}. \]
\end{lemma}
\begin{IEEEproof}
Using Lemma~\ref{lem:SphereBoundGeneral} and \eqref{eq:asymballvol}, we have
\begin{align}
\label{eq:SpherBoundEnt}
    \log_{2^t}\parenv*{V_{2^t,2^m,R_t(\alpha m,m)} }\geq 2^m-\sum_{i=0}^{\alpha m}\binom{m}{i}\geq 2^m-2^{m H_2(\alpha)}=2^m \parenv*{1-2^{-m(1-H_2(\alpha))}}.
\end{align}
Assume to the contrary that $R_t(\alpha m,m)\leq \mu 2^m$ for some $\mu<1-\frac{1}{2^t}$ and infinitely values of $m$. In that case, by \eqref{eq:SpherBoundEnt} and~\eqref{eq:asymballvol},
\begin{align*}
    H_{2^t}(\mu) 2^m &\geq \log_{2^t} \parenv*{V_{2^t,2^m,R_t(r,m)} } \geq 2^m \parenv*{1-2^{-m(1-H_2(\alpha))}}.
\end{align*}
That is, 
\[H_{2^t}(\mu)\geq 1-2^{-m(1-H_2(\alpha))}.\]
Since $\alpha\leq\frac{1}{2}$, we have $H_2(\alpha)<1$, and therefore taking $m\to \infty$ we get $H_{2^t}(\mu)\geq 1$. That is a contradiction as $\mu<1-\frac{1}{2^t}$. This proves that 
\begin{align*}
    \liminf_{m\to\infty} \frac{R_t(\alpha m,m)}{2^m}\geq 1-\frac{1}{2^t}. 
\end{align*}
From the upper bound presented in Lemma~\ref{lem:BinomBound} we have \begin{align*}
    \limsup_{m\to\infty} \frac{R_t(\alpha m,m)}{2^m}\leq 1-\frac{1}{2^t}.
\end{align*}
Combining these two inequalities we have claim.
\end{IEEEproof}

Using the previous two lemmas, we can now state the bound on $R_t(\alpha m,m)$.

\begin{theorem}
\label{th:boundalphalow}
For all $m,t\in\N$, and $0<\alpha\leq\frac{1}{2}$,
\begin{align*}
& \parenv*{1-\frac{1}{2^t}}2^m-\frac{\sqrt{2t(2^t-1)\ln 2}}{2^t}\cdot 2^{\frac{m}{2}(1+H_2(\alpha))}\cdot(1+o(1)) \\
&\qquad \leq R_t(\alpha m,m) \\
&\qquad \leq \parenv*{1-\frac{1}{2^t}}2^{m} - \frac{\sqrt{2^t-1}}{2^t}\cdot \frac{1}{\sqrt{8m\alpha(1-\alpha)}}\cdot 2^{mH_2(\alpha)},
\end{align*}
where $t$ and $\alpha$ are constants.
\end{theorem}
\begin{IEEEproof}
The upper bound follows immediately from~\eqref{eq:asymbinom} and Lemma~\ref{lem:BinomBound}. We turn to prove the lower bound. By Lemma~\ref{lem:SphereBoundGeneral} and~\eqref{eq:asymballvol} again,
\begin{align}
\label{eq:EntIneq}
    2^m H_{2^t}\parenv*{\frac{R_t(\alpha m,m)}{2^m}}
    \geq
    \log_{2^t}\parenv*{V_{2^t,m,R_t(\alpha m,m)}}
    \geq
    2^m-2^{m H_2(\alpha)}.
\end{align}
Since Lemma~\ref{lem:BinomBound} implies $R_t(\alpha m,m)<\parenv{1-\frac{1}{2^t}}{2^m}$, we denote $y\eqdef 1-1/2^t-R_t(\alpha m,m)/2^m>0$. By Lemma~\ref{lem:limitrtlowalpha}, $y=o(1)$. In a similar fashion to the proof of Theorem~\ref{th:lowerbound-fixs}, by~\eqref{eq:enttaylor} we have 
\[H_{2^t}\parenv*{\frac{R_t(\alpha m,m)}{2^m}}= 1-c y^2(1+o(1)),\]
where $c=\frac{2^{2t}}{2t(2^t-1)\ln 2}$. Substituting this back into~\eqref{eq:EntIneq} we get
\[   1-c y^2(1+o(1))\geq 1-2^{m(H_2(\alpha)-1)},\]
and therefore,
\begin{align*}
 y\leq c^{-\frac{1}{2}} 2^{\frac{m}{2}(H_2(\alpha)-1)}  (1+o(1)).
\end{align*}
Since $R_t(\alpha m,m)=(1-1/2^t-y)2^m$, we reach the claimed lower bound.
\end{IEEEproof}

In the region $0<\alpha \leq 1 - \frac{1}{\sqrt{2}}$, we follow a similar procedure to that of~\cite{cohen1992covering}, in order to improve the upper bound of Theorem~\ref{th:boundalphalow}. The following lemma is a sharpening of Lemma~\ref{lem:BinomBound}, requiring more involved work.

\begin{lemma}
\label{lem:ImprovedBinBound}
For all $m,t\in\N$, $2\leq r\leq \frac{m}{2+\sqrt{2}}$, and $m\geq 3$,
\[ R_t(r,m)\leq \parenv*{1-\frac{1}{2^t}}2^m-\frac{\sqrt{2^t-1}}{2^t}\parenv*{1+\sqrt{2}}^{r-1}2^{\frac{m-1}{2}}+\frac{\sqrt{2^t-1}}{2^{t}\sqrt[4]{2} }r\binom{m}{r}.\] 
\end{lemma}

\begin{IEEEproof}
Like the proof of Lemma~\ref{lem:BinomBound}, we proceed by induction on $m$. Throughout this proof we denote the constant $\frac{\sqrt{2^t-1}}{2^t}$ by $c$.  As base cases we shall consider both the case of $m= \ceil{(2+\sqrt{2})r}$ and $r\geq 2$, as well as the case of $r=2$ for all $m$.

Assume that $m= \ceil{(2+\sqrt{2})r}$ and $r\geq 2$. We first observe that
\begin{equation}
    \label{eq:entropy1}
H_2\parenv*{\frac{1}{2+\sqrt{2}}}=\frac{1}{2}+\frac{1}{2+\sqrt{2}}\log_2\parenv*{1+\sqrt{2}}.
\end{equation}
Additionally, by simply monotonicity, as well as~\eqref{eq:asymbinom} and the comment following it,
\begin{equation}
    \label{eq:mrbinom}
    \binom{m}{r}=
    \binom{\ceil{r\parenv{2+\sqrt{2}}}}{r  }\geq  \binom{r\parenv{2+\sqrt{2}}}{r  }
    \geq \frac{\sqrt[4]{2}}{\sqrt{8r } } 2^{r(2+\sqrt{2}) H_2\parenv*{\frac{1}{2+\sqrt{2}}}}.
\end{equation}
We now have the following sequence of inequalities proving the first base case,
\begin{align*}
     R_t(r,m)& \overset{(a)}{\leq} \parenv*{1-\frac{1}{2^t}}2^m \\
     &\overset{(b)}{\leq} \parenv*{1-\frac{1}{2^t}}2^m - c\parenv*{\frac{1}{1+\sqrt{2}}-\frac{\sqrt{r}}{\sqrt{8}}}2^{r\log_2(1+\sqrt{2})+\frac{r}{2}(2+\sqrt{2})}
     \\
     &\overset{(c)}{=}  \parenv*{1-\frac{1}{2^t}}2^m -\frac{c}{1+\sqrt{2}}(1+\sqrt{2})^r 2^{\frac{r}{2}(2+\sqrt{2})}+\frac{cr}{\sqrt{8r}}2^{r(2+\sqrt{2})H_2\parenv*{\frac{1}{2+\sqrt{2}}}}\\
     & \overset{(d)}{\leq}   \parenv*{1-\frac{1}{2^t}}2^m -c(1+\sqrt{2})^{r-1}2^{\frac{m-1}{2}}+\frac{cr}{\sqrt[4]{2}}\binom{m}{r},
\end{align*}
where $(a)$ follows from Lemma~\ref{lem:BinomBound}, $(b)$ follows since for all $r\geq 2$ we have $\frac{1}{1+\sqrt{2}}\leq \frac{\sqrt{r}}{\sqrt{8}}$, $(c)$ follows from~\eqref{eq:entropy1}, and $(d)$ follows since $m= \ceil{(2+\sqrt{2})r}$ as well as by~\eqref{eq:mrbinom}.

We now check that the claim holds for the second base case, where $r=2$. We observe that,
\begin{align*}
    R_t(2,m) &\overset{(a)}{\leq} \sum_{i=2}^{m-1} R_t(1,i) \\
    &\overset{(b)}{\leq} \parenv*{1-\frac{1}{2^t}}\parenv*{\sum_{i=2}^{m-1}2^i}-c\sum_{i=2}^{m-1}(\sqrt{2})^i\\
   & \leq \parenv*{1-\frac{1}{2^t}}2^m -c\sum_{i=2}^{m-1}(\sqrt{2})^i\\
   &=\parenv*{1-\frac{1}{2^t}}2^m - c \parenv*{
   (1+\sqrt{2})2^{\frac{m}{2
   }}-\frac{2}{\sqrt{2}-1}},
\end{align*}
where $(a)$ follows by repeated application of Proposition~\ref{prop:recursive} and the fact that $R_t(2,2)=0$, and $(b)$ follows from Lemma~\ref{lm:upperbound-s=1}. We note that the base case is proved when
\begin{align}
    \label{eq:AnoyingCalc}
    (1+\sqrt{2})2^{\frac{m}{2}}-\frac{2}{\sqrt{2}-1}\geq (1+\sqrt{2})2^{\frac{m-1}{2}}-\frac{1}{\sqrt[4]{2}}\cdot 2\cdot\binom{m}{2},
\end{align}
is satisfied. Indeed, one can easily check that \eqref{eq:AnoyingCalc} holds for all $m\geq \ceil{(2+\sqrt{2})2}=7$.

Having completed the induction base cases, assume the claim holds for $m-1$, i.e., for all $2\leq r\leq \frac{m-1}{2+\sqrt{2}}$. We shall now prove the claim also holds for $m$, and all $2\leq r\leq \frac{m}{2+\sqrt{2}}$. The two extreme cases, i.e., $r=2$, and $m=\ceil{(2+\sqrt{2})r}$, have already been proved in the base cases. For the remaining values of $r$,
\begin{align*}
    R_t(r,m)&\leq R_t(r-1,m-1)+R_t(r,m-1)\\
    & \leq \parenv*{1-\frac{1}{2^t}}2^{m-1}-c(1+\sqrt{2})^{r-2}2^{\frac{m-2}{2}}+\frac{c}{\sqrt[4]{2}}(r-1)\binom{m-1}{r-1}\\
    & \quad + \parenv*{1-\frac{1}{2^t}}2^{m-1}-c(1+\sqrt{2}) ^{r-1}2^{\frac{m-2}{2}}+\frac{c}{\sqrt[4]{2}}r\binom{m-1}{r}\\
    &\leq \parenv*{1-\frac{1}{2^t}}2^{m}-c(1+\sqrt{2})^{r-1}\parenv*{1+\frac{1}{1+\sqrt{2}}}2^{\frac{m-2}{2}}+\frac{cr}{\sqrt[4]{2}}\parenv*{\binom{m-1}{r}+\binom{m-1}{r-1}}\\
    &= \parenv*{1-\frac{1}{2^t}}2^{m}-c(1+\sqrt{2})^{r-1}2^{\frac{m-1}{2}}+\frac{cr}{\sqrt[4]{2}}\binom{m}{r},
\end{align*}
where the first inequality follows from Proposition~\ref{prop:recursive}, and then we use the induction hypothesis.
\end{IEEEproof}

\begin{theorem}
\label{th:boundalphalowbetter}
For all $m,t\in\N$, and $0<\alpha<1-\frac{1}{\sqrt{2}}$,
\[ R_t(\alpha m, m)\leq \parenv*{1-\frac{1}{2^t}}2^m-\frac{\sqrt{2^t-1}}{2^t(2+\sqrt{2})}2^{ m \parenv*{\frac{1}{2}+\alpha \log_2(1+\sqrt{2})}}(1+o(1)),\]
where $t$ and $\alpha$ are constants.
\end{theorem}

\begin{IEEEproof}
By Lemma~\ref{lem:ImprovedBinBound},
\begin{align*}
    R_t(\alpha m,m) &\leq \parenv*{1-\frac{1}{2^t}}2^m-\frac{\sqrt{2^t-1}}{2^t}\parenv*{(1+\sqrt{2}) ^{\alpha m-1}2^{\frac{m-1}{2}}-\frac{\alpha m}{\sqrt[4]{2}}\binom{m}{\alpha m}} \\
    &=\parenv*{1-\frac{1}{2^t}}2^m-\frac{\sqrt{2^t-1}}{2^t(2+\sqrt{2})}2^{ m \parenv*{\frac{1}{2}+\alpha \log_2(1+\sqrt{2})}}+2^{m(H_2(\alpha)+o(1)) }\\
    &=\parenv*{1-\frac{1}{2^t}}2^m-\frac{\sqrt{2^t-1}}{2^t(2+\sqrt{2})}2^{ m \parenv*{\frac{1}{2}+\alpha \log_2(1+\sqrt{2})}}(1+o(1)),
\end{align*}
where we made use of~\eqref{eq:asymbinom}, and the fact that
\[ \frac{1}{2}+\alpha \log_2(1+\sqrt{2}) > H_2(\alpha).\]
\end{IEEEproof}

\begin{table}
\caption{A summary of exact values}
\label{tab:exact}
\begin{center}
\renewcommand{\arraystretch}{2}
\begin{tabular}{c|l|l}
\hline
\hline
$R_t(0,m)$ & $2^m-\ceil*{2^{m-t}}$ & Proposition~\ref{prop:rtzerom} \\ \hline
$R_t(m-2,m)$ & $\min\set*{t,m}+1$ & Proposition~\ref{prop:rtalmostm} \\ \hline
$R_t(m-1,m)$ & $1$ & Proposition~\ref{prop:rtalmostm} \\ \hline
$R_t(m,m)$ & $0$ & Proposition~\ref{prop:rtalmostm}\\
\hline
\hline
\end{tabular}
\end{center}
\end{table}

\begin{table*}
\caption{A summary of the bounds}
\label{tab:summary}
\begin{center}
\renewcommand{\arraystretch}{2}
\begin{tabular}{c|l|l}
\hline
\hline
\multirow{2}{*}{$R_t(r,m)$} & 
$\leq \parenv*{1-\frac{1}{2^t}}2^m-\frac{\sqrt{2^t-1}}{2^t} (1+\sqrt{2})^{r-1}2^{m/2}+O(m^{r-2})$ &
Theorem~\ref{th:upperbound-fixs} \\[2pt]
\cline{2-3}
&
$\geq \parenv*{1-\frac{1}{2^t}}2^m-\frac{\sqrt{2t(2^t-1) \ln 2}}{2^t\sqrt{r!}} m^{r/2}2^{m/2}(1+o(1))$ &
Theorem~\ref{th:lowerbound-fixs}\\[2pt]
\hline
\hline
\multirow{2}{*}{$R_t(m-s,m)$} &
$\leq \frac{t}{(s-2)!}m^{s-2}+O(m^{s-3})$ &
\multirow{2}{*}{Theorem~\ref{th:LB2}}\\[2pt]
\cline{2-2}
&
$\geq \frac{t}{(s-1)!} m^{s-2}+O(m^{s-3}\log(m))$ & \\[2pt]
\hline
\hline
\multirow{5}{*}{$R_t(\alpha m,m)$} &
$\leq \parenv*{1-\frac{1}{2^t}}2^m-\frac{\sqrt{2^t-1}}{2^t(2+\sqrt{2})}2^{ m \parenv*{\frac{1}{2}+\alpha \log_2(1+\sqrt{2})}}(1+o(1))$ &
Theorem~\ref{th:boundalphalowbetter}, assuming $0<\alpha<1-\frac{1}{\sqrt{2}}$ \\[2pt]
\cline{2-3}
&
$\leq \parenv*{1-\frac{1}{2^t}}2^{m} - \frac{\sqrt{2^t-1}}{2^t}\cdot \frac{1}{\sqrt{8m\alpha(1-\alpha)}}\cdot 2^{mH_2(\alpha)}$ &
Theorem~\ref{th:boundalphalow}, assuming $1-\frac{1}{\sqrt{2}}\leq\alpha\leq\frac{1}{2}$ \\[2pt]
\cline{2-3}
&
$\leq t\cdot 4^{H_2(\alpha)}\cdot 2^{mH_2(\alpha)}\cdot (1+o(1))$ &
Theorem~\ref{th:highalpha}, assuming $\frac{1}{2}<\alpha<1$ \\[2pt]
\cline{2-3}
&
$\geq \parenv*{1-\frac{1}{2^t}}2^m-\frac{\sqrt{2t(2^t-1)\ln 2}}{2^t}\cdot 2^{\frac{m}{2}(1+H_2(\alpha))}\cdot(1+o(1))$ &
Theorem~\ref{th:boundalphalow}, assuming $0<\alpha\leq\frac{1}{2}$ \\[2pt]
\cline{2-3}
&
$\geq t\cdot \sqrt{\frac{1-\alpha}{8(\alpha m)^3}}\cdot  2^{mH_2(\alpha)}\cdot(1+o(1))$ &
Theorem~\ref{th:highalpha}, assuming $\frac{1}{2}<\alpha<1$ \\[2pt]
\hline
\hline
\end{tabular}
\end{center}
\end{table*}

\section{Covering Algorithm}
\label{sec:alg}

In this section we describe an algorithm which receives as input a matrix $\bv\in\F_2^{t\times 2^m}$, and returns a codeword matrix $\bc\in\rmc^t(r,m)$ that is no farther away from $\bv$ than the upper bounds described in the previous section, namely
\[ d^{(t)}(\bv,\bc)\leq U_t(r,m),\]
where $U_t(r,m)$ is any upper bound on $R_t(r,m)$ from Table~\ref{tab:summary}. We call this a \emph{covering algorithm}, and it may be thought of as the analogue to a decoding algorithm for an error-correcting code.

To motivate our study of a covering algorithm, we recall the motivating example described in~\cite{elimelech2020generalized}. We look at linear data querying schemes, the most prominent example of which is private information retrieval (PIR), in which the user queries a database by linear combinations. We think of the database as a sequence of elements $\ox=(x_1,\dots,x_m)\in \F^m_{q^\ell}$. The user may query the contents of the database by providing $\os=(s_1,\dots,s_m)\in \F^m_q$, and getting in response the linear combination $\os\cdot\ox=\sum_{i=1}^m s_i x_i$. The access complexity in such protocols is the number of database items that need to be read in order to compute the desired linear combination. In a straightforward implementation, the access complexity is the number of non-zero coefficients in $s_1,\dots,s_m$. Thus, in a typical PIR scheme, which selects random coefficients, the expected fraction of non-zero coefficients is $1-\frac{1}{q}$, resulting in a prohibitively high access complexity.

In order to reduce the access complexity one may pre-compute and store some linear combinations of data elements. If the original database is $\ox=(x_1,\dots,x_m)\in\F_{q^\ell}^m$, the linear combinations $\oh_1\cdot\ox,\oh_2\cdot\ox,\dots,\oh_n\cdot\ox$ are pre-computed and stored instead of the original database $\ox$, where $\oh_1,\dots,\oh_n\in \F_q^m$. Assume now that the database receives a query given by $\os\in\F_q^m$. If we can find $r\leq m$ vectors $\oh_{i_1},\dots, \oh_{i_r}$ such that $\os\in \spn{\oh_{i_1},\dots , \oh_{i_r}}$, then we may answer the query by accessing the $r$ pre-computed linear combinations $\oh_{i_1}\cdot\ox,\dots , \oh_{i_r}\cdot\ox$, instead of accessing all the $m$ elements in the database, $x_1,\dots,x_m$. Considering the vectors $\oh_1,\dots,\oh_n$ as the columns of a parity-check matrix $H$ of an  $[n,n-m]_q$ linear code $C$, Definition~\ref{def:rt1} guarantees that $r\leq R_1(C)$ such vectors may always be found. Thus, by storing the $n$ pre-computed linear combinations instead of the original database, we increased the storage, but we reduced the access complexity since we need to access at most $R_1(C)$ elements of the database. As an additional step, assume the database does not answer queries individually, but instead groups together $t$ queries given by $\os_1,\dots,\os_t\in\F_q^m$. We now need the $r$ vectors $\oh_{i_1},\dots,\oh_{i_r}$ to satisfy $\os_1,\dots,\os_t\in\spn{\oh_{i_1},\dots,\oh_{i_r}}$ in order to answer the queries. By Definition~\ref{def:rt1}, $r\leq R_t(C)$ such vectors exist, and by Lemma~\ref{lem:subadd}, $R_t(C)\leq t R_1(C)$. Thus, by delaying the answers to queries, namely, increasing the latency, we have further reduced the access complexity from $t R_1(C)$ (the access complexity of treating $t$ queries individually) to $R_t(C)$.

We translate this problem into a more convenient form. Let us write the vectors $\os_1,\dots,\os_t$ as rows of a matrix $\bs\in\F_q^{t\times m}$. Since the parity-check matrix of $C$ is a full-rank matrix, $H\in\F_q^{m\times n}$, by solving a set of linear equations we can efficiently find a matrix $\bv\in\F_q^{t\times n}$ such that $H\bv^\T=\bs^\T$. We would now like to solve the following task: Given $\bv\in\F_q^{t\times n}$, find $\bc\in C^t$ such that $d^{(t)}(\bv,\bc)\leq r$. We observe that by finding such $\bc$, since $H(\bv-\bc)^\T=\bs^\T$, the rows of $\bv-\bc$ describe linear combinations of the columns of $H$ that both result in $\os_1,\dots,\os_t$, and use no more than $r$ columns. Ideally, we would like to choose $r=R_t(C)$.

We call such an algorithm a \emph{$t$-covering algorithm for $C$, with radius $r$}. It bears a resemblance to a decoding algorithm for an error-correcting code, however some crucial differences are to be noted. To guarantee unique decoding, standard decoding algorithms assume the input is a point in the space that is no more than $\floor{\frac{d-1}{2}}$ away from a codeword, where $d$ is the minimum distance of the code. The covering algorithm may receive as input \emph{any} point in the metric space. Additionally, the decoding algorithm returns the closest (and only) codeword within radius of $\floor{\frac{d-1}{2}}$ from the input point. In contrast, the covering algorithm may return \emph{any} codeword whose distance from the input as it most $r$, and not necessarily the closest codeword. Thus, the covering algorithm discussed here does not perform maximum-likelihood decoding.

As we saw in Section \ref{sec:bounds}, computing the the generalized covering radii of Reed-Muller codes is a difficult task in general. Even for the case of $t=1$, and despite having been studied for decades, the covering radius of Reed-Muller codes is still not fully known. Thus, finding an efficient $t$-covering algorithm for $\rmc(r,m)$, with radius $R_t(r,m)$, poses a great challenge, if only for the fact that $R_t(r,m)$ is unknown in general. An inefficient, brute-force implementation of such an algorithm is trivial, yet, uninteresting.

Instead, in what follows, we devise an efficient $t$-covering algorithm for $\rmc(r,m)$, with radius $U_t(r,m)$, where $U_t(r,m)$ is any of the upper bounds on $R_t(r,m)$ found in this paper, and summarized in Table~\ref{tab:summary}. Our approach stems from the fact that all the bounds in Table~\ref{tab:summary} are derived recursively using the $(u,u+v)$ construction (Proposition~\ref{prop:recursive}) and subadditivity (Lemma~\ref{lem:subadd}), as well as simple base cases.

\begin{algorithm}[t]
\label{alg:1}
\DontPrintSemicolon
\SetKwInOut{Input}{Input}
\SetKwInOut{Output}{Output}
\SetKwFunction{recur}{recursive}
\SetKwFunction{subadd}{subadditive}
\SetKwFunction{cover}{cover}
\SetKwProg{Fn}{Function}{}{}
\Fn{\recur{$\bv$, $r$}}{
\Input{%
$\bv\in\F_2^{t\times 2^m}$, $r\in\N$, $1\leq r\leq m$
}
\tcp{Check edge cases}
\lIf {$r=m$}{
 \Return $\bv$
}
\lIf {$r=1$}{
\Return $\argmin_{\bc\in\rmc(1,m)^t} d^{(t)}(\bv,\bc)$
}
\tcp{Use the $(u,u+v)$ recursion}
Let $\bv_1,\bv_2\in \F_2^{t\times 2^{m-1}}$ s.t.~$\bv=(\bv_1,\bv_2)$\;
$\bc_1\gets \recur(\bv_1,r)$\;
$\bc_2\gets \recur(\bv_2-\bc_1,r-1)$\;
\Return $(\bc_1,\bc_1+\bc_2)$\;
}
\Fn{\subadd{$\bv$, $r$}}{
\Input{%
$\bv\in\F_2^{t\times 2^m}$, $r\in\N$, $1\leq r\leq m$
}
\tcp{Use subadditivity}
Let $\ov_i$ be the $i$-th row of $\bv$\;
\ForAll{$i\in[t]$}{
$\oc_i\gets\recur(\ov_i,r)$
}
\Return $(\oc_1^\T,\dots,\oc_t^\T)^\T$ \;
%\Return $\begin{pmatrix} \oc_1 \\ \vdots \\ \oc_t\end{pmatrix}$\;
}
\Fn{\cover{$\bv$, $r$}}{
\Input{%
$\bv\in\F_2^{t\times 2^m}$, $r\in\N$, $1\leq r\leq m$
}
$\bc_{\min}\gets \recur(\bv,r)$\;
$\bc'_{\min}\gets \subadd(\bv,r)$\;
\Return $\argmin_{\bc\in\set{\bc_{\min},\bc'_{\min}}} d^{(t)}(\bv,\bc)$
}
\caption{A $t$-covering algorithm for $\rmc(r,m)$ with radius $U_t(r,m)$}
 
\end{algorithm}

\begin{theorem}
\label{th:alg}
For any $t,r,m\in\N$, $r\leq m$, and any $\bv\in\F_2^{t\times 2^m}$, running $\bc=\cover(\bv,r)$, from Algorithm~\ref{alg:1}, produces $\bc\in\rmc(r,m)^t$ such that $d^{(t)}(\bv,\bc)\leq U_t(r,m)$. Additionally, its run-time complexity is $O(t 2^t (2^{t+1})^{m+1} (2^{t+1}-1)^{-r}+tm2^m)$.
\end{theorem}

\begin{IEEEproof}
The algorithm clearly stops since, during the recursive calls, either $r$ or $m$ strictly decrease, and the base cases of $r=1$ and $r=m$ are eventually reached. The returned $\bc$ is clearly a codeword, stemming from the base cases and the $(u,u+v)$ structure of Reed-Muller codes. Finally, $d^{(t)}(\bv,\bc)\leq U_t(r,m)$ due to Proposition~\ref{prop:recursive}, Lemma~\ref{lem:subadd}, and the fact that all the bounds in Table~\ref{tab:summary} are relaxations of both (including Theorem~\ref{th:LB2} which is based on a result from~\cite{cohen1992covering}).

We move on to the analysis of the run-time complexity. We first analyze $\recur(\bv,r)$, whose running time we denote by $T(t,r,m)$. We contend that for some constant $c\in\N$, 
\[ T(t,r,m)\leq f(t,r,m) \eqdef c\parenv*{t 2^t (2^{t+1})^{m+1} (2^{t+1}-1)^{-r}+tm2^m}.\]
This proof is by induction. For the first simple base case of $r=m$ we have $T(t,m,m)=c'$, a constant, and indeed
\[ T(t,m,m) =c' \leq c\parenv*{t 2^t (2^{t+1})^{m+1} (2^{t+1}-1)^{-m}+tm2^m}=f(t,m,m),\]
for a proper choice of $c$. Next, we check the base case $r=1$. In this case, a brute-force distance measurement is performed between $\bv$ and the codewords of $\rmc(1,m)^t$. Each codeword is a $t\times 2^m$ matrix, and we have a total of $\abs{\rmc(1,m)^t}=2^{(m+1)t}$ such codewords. Thus, for some constant $c'$,
\[ T(t,1,m) = c'\cdot t2^m\cdot 2^{(m+1)t} \leq c\parenv*{t 2^t (2^{t+1})^{m+1} (2^{t+1}-1)^{-1}+tm2^m}=f(t,1,m),\]
for any $c\geq c'$. Moving on to the main recursion, assume the claim holds for $T(t,r,m-1)$, for all $1\leq r\leq m-1$, and we prove it also holds for $T(t,r,m)$ for all $1\leq r\leq m$. If $r=1$ or $r=m$, we have a base case which we have already proved. Otherwise, the algorithm manipulates a $t\times 2^m$ matrix and runs two recursive instances. Hence, for some constant $c'$, and after choosing any $c\geq c'$, we have
\begin{align*}
    T(t,r,m) & = c' t 2^m + T(t,r-1,m-1)+ T(t,r,m-1) \\
    &\leq ct2^m + c\parenv*{t 2^t (2^{t+1})^{m} (2^{t+1}-1)^{-r+1}+t(m-1)2^{m-1}} + c\parenv*{t 2^t (2^{t+1})^{m} (2^{t+1}-1)^{-r}+t(m-1)2^{m-1}} \\
    & = c\parenv*{t 2^t (2^{t+1})^{m+1} (2^{t+1}-1)^{-r}+tm2^m} \\
    &= f(t,r,m).
\end{align*}
This completes the induction. To complete the proof as well, we note that the complexity of $\subadd(\bv,r)$ is always subsumed by the complexity of $\recur(\bv,r)$.
\end{IEEEproof}

As in the previous section, we analyze three asymptotic regimes for $r$ and $m$:

\begin{corollary}
Let $t\in\N$ be a constant, let $n=2^m$ be the length of the code $\rmc(r,m)$, and denote $\beta\eqdef \log_2\sqrt[t+1]{2^{t+1}-1}$. Then the run-time complexity of Algorithm~\ref{alg:1} is:
\begin{itemize}
    \item 
    $O(n^{t+1})$ when $r$ is constant.
    \item
    $O(n^{(t+1)(1-\alpha\beta)})$ when $r=\alpha m$, and $0<\alpha<\frac{t}{(t+1)\beta}$ is a constant.
    \item
    $O(n\log n)$ when $r=m-s$, $s$ is constant, or when $r=\alpha m$, and $\frac{t}{(t+1)\beta}\leq \alpha<1$ is a constant.
\end{itemize}
\end{corollary}

\begin{IEEEproof}
This is a straightforward application of Theorem~\ref{th:alg}. The $\frac{t}{(t+1)\beta}$ cutoff point stems from the fact that the complexity is in fact $O(n^{(t+1)(1-\alpha\beta)}+n\log n)$. Thus, for $\alpha<\frac{t}{(t+1)\beta}$, we have that $(t+1)(1-\alpha\beta)>1$, and $n^{(t+1)(1-\alpha\beta)}$ dominates the complexity. However, when $\alpha\geq\frac{t}{(t+1)\beta}$, we have that $(t+1)(1-\alpha\beta)\leq 1$ and $n\log n$ dominates the complexity.
\end{IEEEproof}

\section{Conclusion}
\label{sec:conc}

In this work, we studied the generalized covering radii of Reed-Muller codes, $R_t(r,m)$. In some simple cases we found the exact generalized covering radii (see Table~\ref{tab:exact}). For most other cases we found lower and upper bounds on the generalized covering radii (see Table~\ref{tab:summary}). These bounds were found in three asymptotic regimes: $r$ constant, $m-r$ constant, and $r/m$ constant. We also constructed a $t$-covering algorithm with radius no worse than the upper bounds that we found (see Algorithm~\ref{alg:1}). We analyzed the algorithm's run-time complexity and showed it is polynomial in the code parameters.

\begin{figure}[t]
\centering
\begin{overpic}[scale=0.4]
    {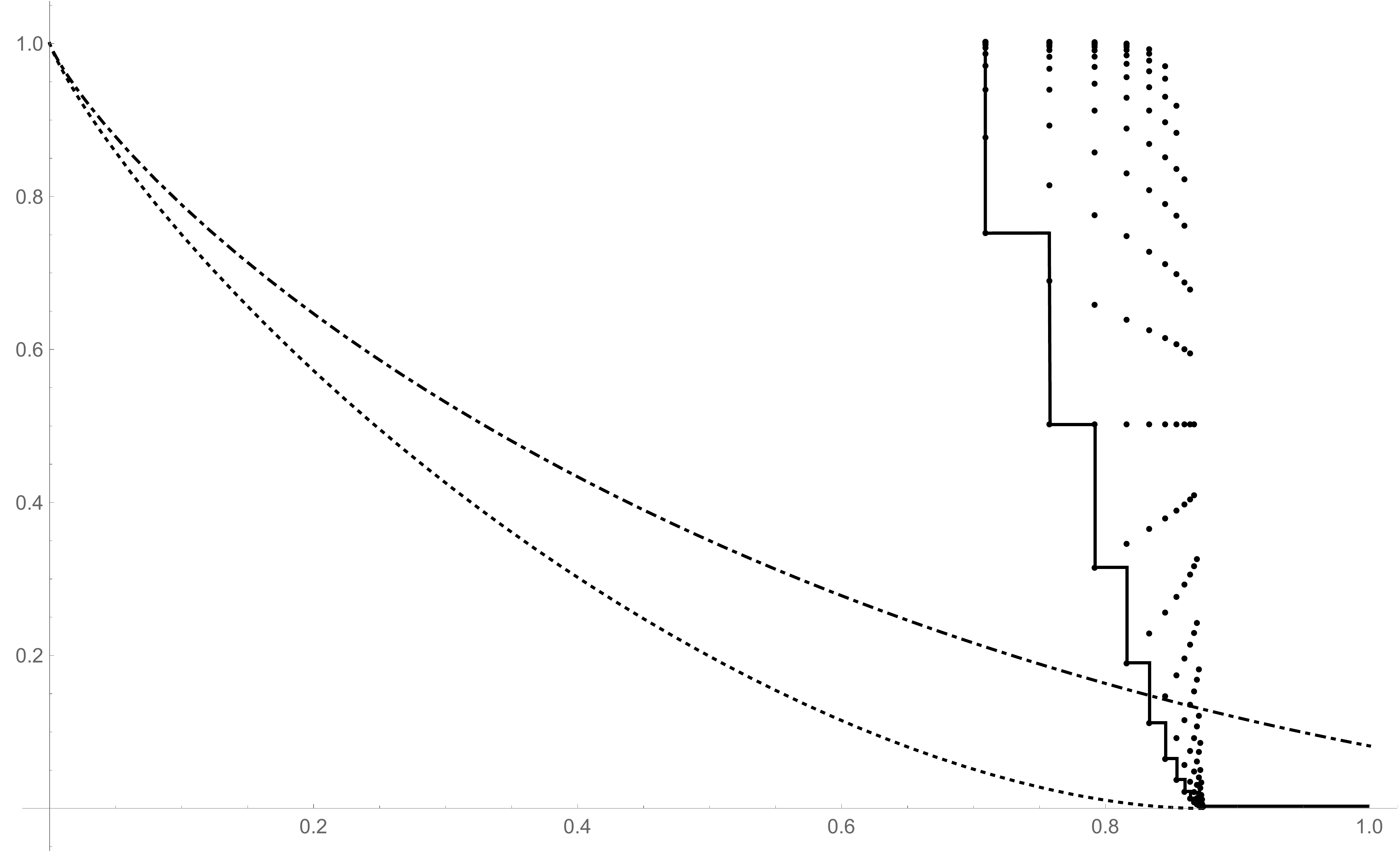}
    \put(-4,35){\begin{turn}{90}$\kappa_3(\rho,2)$\end{turn}}
    \put(55,-1){$\rho$}
    \put(23,30){(a)}
    \put(48,26){(b)}
    \put(76,18.8){(c)}
\end{overpic}
\caption{
A comparison of the bounds on $\kappa_3(\rho,2)$: (a) the ball-covering lower bound \cite[Prop.~12]{elimelech2020generalized}, (b) the general upper bound \cite[Prop.~14]{elimelech2020generalized}, and (c) our upper bound obtained from the upper bound on the $t$-th generalized covering radius of Reed-Muller codes.
}
\label{fig:comp}
\end{figure}

\begin{figure}[t]
\centering
\begin{overpic}[scale=0.54]
    {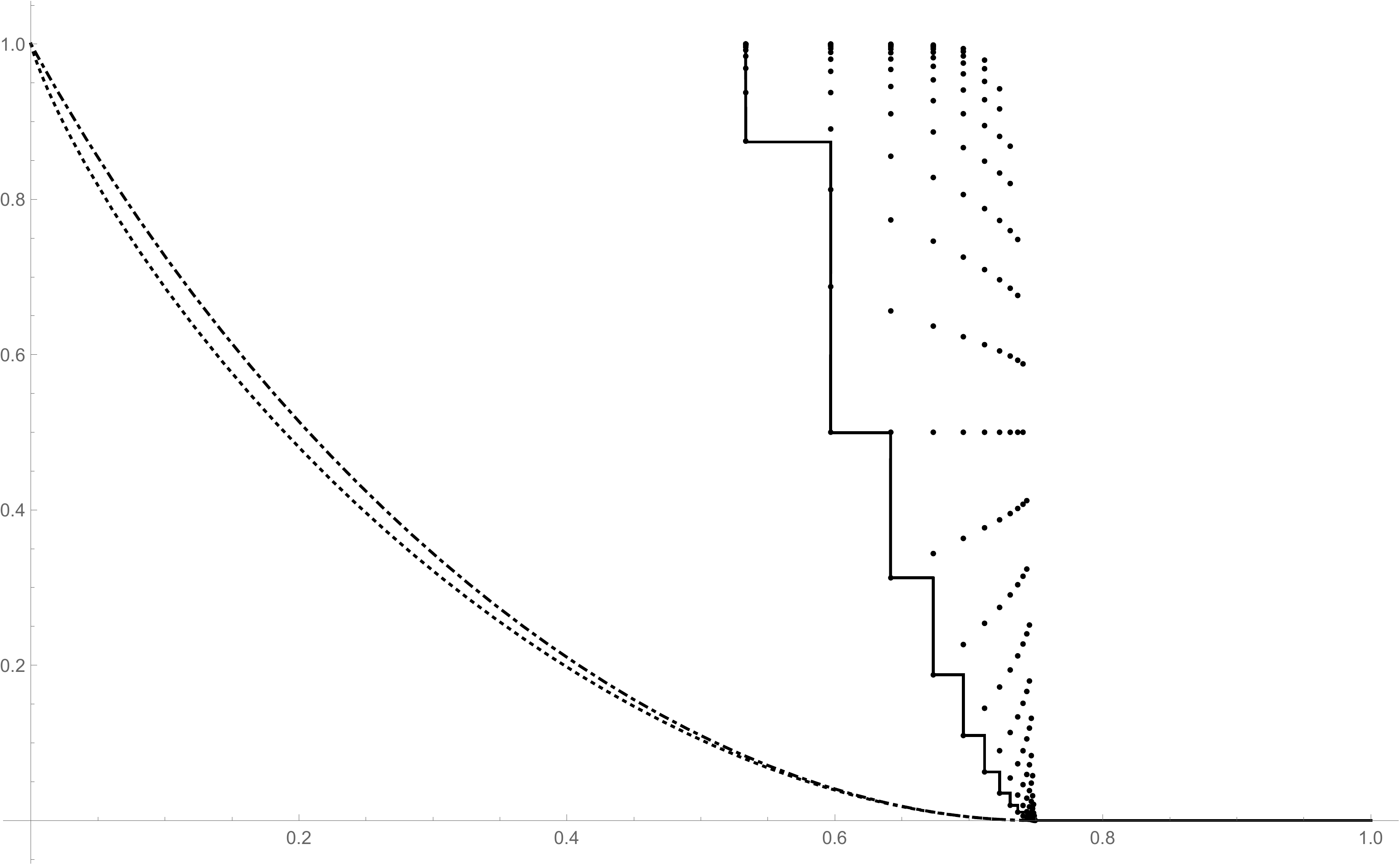}
    \put(-4,35){\begin{turn}{90}$\kappa_2(\rho,2)$\end{turn}}
    \put(55,-1){$\rho$}
    \put(17,30){(a)}
    \put(28,26){(b)}
    \put(63,15){(c)}
\end{overpic}
\caption{
A comparison of the bounds on $\kappa_2(\rho,2)$: (a) the ball-covering lower bound \cite[Prop.~12]{elimelech2020generalized}, (b) the improved upper bound \cite[Thm.~22]{elimelech2020generalized}, and (c) our upper bound obtained from the upper bound on the $t$-th generalized covering radius of Reed-Muller codes.
}
\label{fig:comp2}
\end{figure}
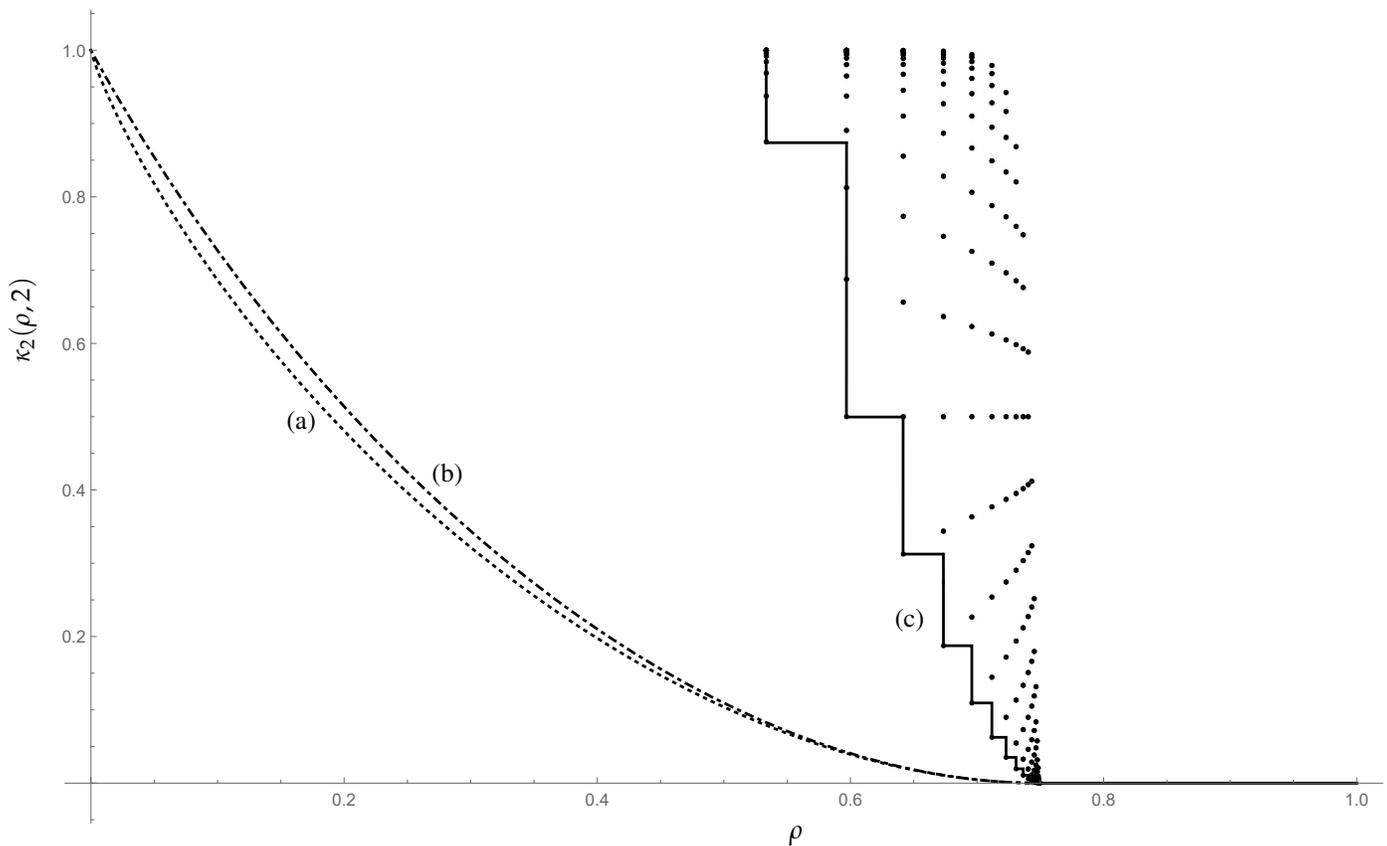

We remark that our upper bounds on the covering radii of Reed-Muller codes may also be used for the study of the asymptotic behaviour of generalized covering radii of linear codes in general. Given the parameters $t\in \N$, $\rho\in [0,1]$ and a prime power $q$, the asymptotic minimal rate of a code over $\F_q$ with a normalized $t$-th generalized covering radius of no more than $\rho$, is denoted by $\kappa_t(\rho, q)$. Since the $t$-th generalized covering radius of a direct sum of codes is the sum of the $t$-th generalized covering radii of its component codes (see \cite[Prop.~25]{elimelech2020generalized}), an $[n,k]_q$ linear code with $t$-th generalized covering radius of $r$ immediately creates an infinite family of codes with rate $\frac{k}{n}$ and normalized $t$-th generalized covering radius $\frac{r}{n}$. It then follows that $\kappa_t(r/n,q)\leq k/n$. By the monotonicity of $\kappa_t(\rho,q)$ in $\rho$, this upper bound holds for all $\rho\geq r/n$. Thus, our upper bounds on the generalized covering radii of Reed-Muller codes (denoted by $U_t(r,m)$) give the following upper bounds:
\begin{equation}
    \label{eq:DiscreteBound}
    \kappa_t(\rho,2)\leq \frac{\dim\parenv*{\rmc(r,m)}}{2^m} \text{ for all } \rho \geq \frac{U_t(r,m)}{2^m} 
    \end{equation}
In Figure \ref{fig:comp}, the bound obtained by applying \eqref{eq:DiscreteBound} in the range $2\leq m \leq 20$, $1\leq r \leq m$ in the case where $t=3$ is presented. Each pair $(r,m)$, results in a point depicted in the graph. We observe that some of the points obtained in this way improve upon the upper bound from \cite[Prop.~14]{elimelech2020generalized},
\begin{equation}
\label{eq:oldsimpleub}
\kappa_t(\rho,q) \leq 1-H_q\parenv*{\frac{\rho}{t}},
\end{equation}
where $H_q(\cdot)$ is the $q$-ary entropy function. A similar comparison, with $t=2$, is shown in Figure~\ref{fig:comp2}. However, specifically for $t=2$, the upper bound of~\cite[Theorem~22]{elimelech2020generalized} is stronger than~\eqref{eq:oldsimpleub}, and so the bound of~\eqref{eq:DiscreteBound} offers no improvement.

We would like to mention a couple of interesting open questions pertaining to the results of this paper. We first observe that, apart from the base cases, our bounds are obtained using the $(u,u+v)$ recursion, and subadditivity. We suspect that for improved bounds, a different approach may be needed, perhaps an approach that exploits the unique geometric and combinatorial properties of Reed-Muller codes.

Another open problem concerns Algorithm~\ref{alg:1}. The edge case of $\rmc(1,m)$ is solved in the algorithm using a brute-force approach: the distance between the input, $\bv$, and the codewords of $\rmc(1,m)^t$ is measured exhaustively and naively. However, for $t=1$, the codewords of $\rmc(1,m)$ form a Sylvester-type Hadamard matrix and its complement. Thus, by using the Walsh-Hadamard transform, an efficient measurement of the distance from $\bv$ to the codewords of $\rmc(1,m)$ is possible in $O(n\log n)$ time, instead of the $\Theta(n^2)$ of a naive implementation, where $n=2^m$ is the code length. Whether a similar approach can improve Algorithm~\ref{alg:1} is still unknown.

Finally, and more generally, it is known that the generalized covering radii are monotone non-decreasing in $t$. Thus, any improvement in the bounds on $R_t(r,m)$ may perhaps bring about an improvement in the bounds on the (regular) covering radius of Reed-Muller codes, $R_1(r,m)$. These problems, and others, are left for future research.

\bibliographystyle{IEEEtranS}
%\bibliography{allbib}
\bibliography{Biblio_Commented}

\end{document}